\documentstyle[12pt]{article}
\def\ep{\epsilon}

\def\Tr{{\rm \, Tr \,\!}  }
\def\Det  {{\rm \,\! Det \,\!} }

\def\o{\over}

\def\cdots {\cdot\cdot\cdot}

\def\ni       {\noindent}
\def\lb       {\left( }
\def\rb       {\right) }
\def\lmb      {\left\{ }

\def\lbb     {\left[ }
\def\rbb      {\right] }
\def\comma      { \, , }
\def\period     { \, . }
\def\semiket#1  { \, #1 \, \rangle \, }
\def\del        {  \partial  }

\def\half       {  {1\over 2}  }
\def\abs#1      {  \, \vert #1 \vert \,   }
\def\Im#1    { \, {\rm Im } \, #1  }
\def\Re#1    { \, {\rm Re}  \, #1  }

\def\binom#1#2 { \vecii{ {}_{#1} }{\raisebox{.5ex}{$ {}^{#2} $}} }
\def\sqbinom#1#2 { \Bigl(\begin{array}{c} {}_{#1}
                       \\ \raisebox{.5ex}{${}^{#2}$} \end{array}\Bigr)^2  }
\def\simg    { \ \raisebox{-.5ex}{$ \stackrel{>}{\sim} $} \ }
\def\siml    { \ \raisebox{-.5ex}{$ \stackrel{<}{\sim} $} \ }

\def\bfZ     { {\bf Z}}

\def\sql    {\sqrt{\lambda}}
\def\r12    {\frac{r_1}{r_2}}
\def\calD  {{\cal D}}
\def\calO  {{\cal O}}
\def\sigbar {\bar{\sigma}}

\def\tili   {\tilde{i}}

\def\tilw  {{\tilde{w}}}
\def\tilu  {{\tilde{u}}}

\def\sighat { {\hat{\sigma}} }
\def\Dhat   { {\hat{D}} }
\def\omhat  { {\hat{\omega}} }
\def\Delhat  { {\hat{\Delta}} }

\def\cg      {c_\gamma}
\def\sg      {s_\gamma}
\def\cp      {c_\psi}
\def\sp      {s_\psi}
\def\vecii#1#2      {  { #1 \choose #2 }  }
\def\veciii#1#2#3   {  \left(\begin{array}{c}#1\\#2\\#3\end{array}\right)  }
\def\matrixii#1#2#3#4            {  \Biggl( \begin{array}{cc}#1&#2\\#3&#4
                                       \end{array} \Biggr) }
\def\matrixiii#1#2#3#4#5#6#7#8#9 {  \left(\begin{array}{ccc}#1&#2&#3\\
                                     #4&#5&#6\\#7&#8&#9\end{array}\right)  }
\def\eqb         {  \begin{eqnarray}  }
\def\eqe           {  \end{eqnarray}  }
\def\nn               {  \nonumber  }
\def\sectionnumbering { \setcounter{equation}{0}
         \renewcommand{\theequation}{\arabic{section}.\arabic{equation}}}
\def\appendixnumbering { \setcounter{equation}{0}
         \renewcommand{\theequation}{\Alph{section}.\arabic{equation}}}
\def\msection#1{ \addtocounter{section}{1} \setcounter{subsection}{0}
  \sectionnumbering
   \par \bigskip
      \par \bigskip \noindent
   {\bf \arabic{section} \quad  #1 }
    \par \bigskip}
\def\appsection#1{\addtocounter{section}{1} \setcounter{subsection}{0}
                 \appendixnumbering
      \par \bigskip \noindent
   {\bf Appendix : \quad  #1 }
    \par \bigskip}
\def\msubsection#1{\addtocounter{subsection}{1}
      \par \noindent  {\normalsize\it
      \arabic{section}.\arabic{subsection} \quad #1  }
   \par \medskip }

\def\csectionast#1    { \begin{center}
    {\large\bf #1  }   \end{center} \par \bigskip}
\def\titleandfile#1#2   {  \begin{center}{\large\bf #1}\end{center}
                            \par\begin{flushright} #2 \end{flushright}  }

\renewcommand{\thefootnote}{\fnsymbol{footnote}}

\begin{document}
%
\def\papertitlepage{\baselineskip 3.5ex \thispagestyle{empty}}
\def\preprinumber#1#2#3#4#5{\hfill \begin{minipage}{4.2cm}  #1
              \par\noindent #2
              \par\noindent #3
              \par\noindent #4
              \par\noindent #5
             \end{minipage}}
\renewcommand{\thefootnote}{\fnsymbol{footnote}}
%
%
\papertitlepage \setcounter{page}{0}
\preprinumber{EPHOU-05-002}{KEK-TH-1004}{Imperial/TP/050402}{UTHEP-502}{hep-th/0504123}
\baselineskip 0.8cm \vspace*{1.5cm}
\begin{center}
{\large\bf Quantum fluctuations of rotating strings in $AdS_{5}
\times S^{5}$}
\end{center}
\vskip 4ex \baselineskip 1.0cm
\begin{center}
  { Hiroyuki ~Fuji\footnote[2]{\tt
  fuji@particle.sci.hokudai.ac.jp } } \\
 \vskip -1ex
    {\it Department of Physics, Hokkaido University} \\
 \vskip -2ex
    {\it Sapporo, 060-0810, Japan}\\
 \vskip -1ex
    {\it High Energy Accelerator Research Organization (KEK)} \\
 \vskip -2ex
    {\it Tsukuba, Ibaraki 305-0801, Japan} \\
 \vskip 2ex
    { Yuji  ~Satoh\footnote[3]{\tt y.satoh@imperial.ac.uk}}  \\
 \vskip -1ex
    {\it Blackett Laboratory, Imperial College} \\
 \vskip -2ex
    {\it London, SW7 2BZ, U.K.} \\
 \vskip -1ex
    {\it Institute of Physics, University of Tsukuba} \\
 \vskip -2ex
    {\it Tsukuba, Ibaraki 305-8571, Japan}

\end{center}
\vskip 7ex
%
\baselineskip=3.5ex
\begin{center} {\bf Abstract} \end{center}
We discuss quantum fluctuations of a class of rotating strings in
$AdS_5 \times S^5$. In particular, we develop a systematic method
to compute the one-loop sigma-model effective actions in closed
forms as expansions for large spins. As examples, we explicitly
evaluate the leading terms for the constant radii strings in the
$SO(6)$ sector with two equal spins, the $SU(2)$ sector, and the
$SL(2)$ sector. We also obtain the leading quantum corrections to
the space-time energy for these sectors.

\vskip 2ex
%
%
%
%
%
\vspace*{\fill} \ni April 2005
\newpage
\renewcommand{\thefootnote}{\arabic{footnote}}
\setcounter{footnote}{0} \setcounter{section}{0} \baselineskip =
3.3ex
\pagestyle{plain}

%
\msection{Introduction}

To better understand dynamical aspects of the AdS/CFT
correspondence,
 one may need studies beyond the BPS sectors.
 An important step toward this direction was made
 in \cite{BMN}. Rotating strings in  $AdS_{5} \times S^{5}$ provide
its generalizations, where deeply non-BPS sectors can be probed
 \cite{GKP}-\cite{EMZ}.  (For a review, see \cite{Tseytlin,Beisert}.)
 In particular, for a certain class of rotating strings,
 one can find an exact agreement between unprotected quantities
 on the string and the gauge theory sides,
 which is not necessarily guaranteed by the AdS/CFT correspondence.
 Moreover, the correspondence has been extended, in a
 unified manner, to that of effective theories or general
 solutions \cite{Kruczenski}-\cite{BKSZ}.

A clue for understanding this agreement may be the integrable
structures on both sides (see, e.g., \cite{MZ,AS},
\cite{KMMZ}-\cite{ART}). A role of a certain asymptotic
(``nearly'') BPS condition has also been pointed out \cite{MMT}.
For large spins, the world-sheets of the rotating strings form
nearly null surfaces and the strings become effectively
tensionless \cite{MMT,Mikhailov}, which intuitively means that the
constituents of the strings appear to be free \cite{tensionless}.
However, we do not yet know why we find the exact agreement, and
why it starts to break down at a certain level \cite{SS}.

In this development, the analysis on the string side tends to be
classical because of difficulties in the quantization, despite
that information of the quantized strings is necessary to complete
the correspondence. This contrasts with detailed quantum analysis
on the gauge theory side \cite{Beisert}. As for the quantum
aspects of the rotating strings,\footnote{For related works, see,
e.g., \cite{1loop}-\cite{quntBA}.}
 the one-loop sigma-model
fluctuations and their stability have been studied for the
``constant radii'' strings \cite{FT1,ART,FT2}. Based on these
results, the one-loop corrections to the space-time energy have
been studied numerically for the $SO(6)$ and the $SU(2)$ sectors
with two equal spins \cite{FPT}, and for the $SL(2)$ sector
\cite{PTT}. Furthermore, the leading correction for the $SL(2)$
sector has been matched in a closed form with the finite size
correction of the anomalous dimension on the gauge theory side
\cite{BTZ}. This result can also be extrapolated to the $SU(2)$
sector (with two equal spins), up to subtleties of instability.

In this paper, we discuss quantum fluctuations of the rotating
strings in $AdS_5 \times S^5$. In particular, we develop a
systematic method to compute the one-loop sigma-model effective
actions and the corrections to the space-time energy in such
backgrounds, so that they are obtained in closed forms as
expansions for large spins. As examples, we consider the constant
radii strings in the $SO(6)$ sector with two equal spins, the
$SU(2)$ sector, and the $SL(2)$ sector, and explicitly evaluate
the leading terms in the expansion. We note that it is in
principle possible to carry out the expansion up to any given
order. The asymptotic BPS condition and the effective tensionless
limit seem to be characteristic of the correspondence and useful
for understanding the string in $AdS_{5} \times S^{5}$ itself.
However, their consequences for the quantum string have not been
investigated well. Through concrete computations, we can see how
they work to make the quantum corrections subleading for large
spins.

The organization of this paper is as follows. In section 2, we
summarize necessary ingredients to discuss the one-loop
fluctuations of strings in $AdS_5 \times S^5$. In section 3, we
discuss the quadratic fluctuations of the constant radii strings
in the $SO(6)$ sector with two equal spins. We evaluate the
fluctuation operators in rotated functional bases, so that the
expansion for large spins becomes well-defined. In the course, we
obtain the fermionic fluctuation operator for the generic $SO(6)$
sector. In section 4, we develop a large $J$ (total spin)
expansion of the one-loop effective action. We explicitly evaluate
the leading term for large $J$ (up to and including $\calO(1/J)$)
 in a closed form. Essentially the
same procedures are applied to other sectors in the following
sections. We briefly summarize the results for the $SU(2)$ sector
in section 5, and for the $SL(2)$ sector in section 6. At the
leading order, all the one-loop effective actions take a universal
form, which is proportional to a geometric invariant. In section
7, from the one-loop effective actions, we read off the
corrections to the space-time energy up to and including
$\calO(1/J^2)$. Comparing the results with the finite size
corrections to the anomalous dimension on the gauge theory side
\cite{BTZ}-\cite{HLPS}, we find that the dependence on the winding
numbers and the filling fractions agree with that of the
``non-anomalous'' (zero-mode) part on the gauge theory side.
Relation to the earlier results in the literature is also
discussed. We conclude in section 8. In the appendix, we summarize
how to evaluate a constant which appears in the expression of the
one-loop effective actions.

%
\msection{Preliminaries}
We consider one-loop sigma-model fluctuations of a certain class
of rotating strings in type IIB theroy. Here, we summarize our
notation and ingredients used in the following sections.

\par\bigskip\ni
{\it Coordinates}
\par\medskip

The metric of $AdS_5 \times S^5$ takes a form \eqb
   ds^{2} &=& G_{\mu\nu} dx^{\mu} dx^{\nu} = ds^{2}_{AdS_{5}} + ds^{2}_{S^{5}}
      \comma \nn \\
    -ds^{2}_{AdS_{5}}&=&
       d \rho^2 - \cosh^2\rho \ dt^2 + \sinh^2\rho \ (d \theta^2 +
      \cos^2 \theta \, d \phi^2_4 + \sin^2 \theta \, d\phi_5^2) \comma
        \label{AdS5S5} \\
    -ds^2_{S^5}
       &=& d\gamma^2 + \cos^2\gamma\, d\phi_3^2 +\sin^2\gamma\
     (d\psi^2 +
          \cos^2\psi\, d\phi_1^2+ \sin^2\psi\, d\phi_2^2) \period \nn
\eqe To express the rotating string solutions, it is useful to
introduce complex variables $Z_r$ $(r= 0, ..., 5)$, so that
$AdS_5$ and $S^5$ are expressed as hypersufaces in flat spaces,
 $ |Z_0|^2 - |Z_4|^2 - |Z_5|^2  = 1$ and
 $ |Z_1|^2 + |Z_2|^2 + |Z_3|^2 = 1 $,
respectively. In terms of the above coordinates, one can set \eqb
   Z_r = a_r e^{i \phi_r} \comma \label{Zr}
\eqe with $ \phi_0 = t $ and \eqb
   \begin{array}{lll}
     a_1 = \sin \gamma \, \cos \psi \comma & a_2 = \sin   \gamma \, \sin \psi \comma &
     a_3 = \cos  \gamma \comma \\
     a_4 = \sinh \rho \, \cos \theta \comma & a_5 = \sinh \rho \, \sin \theta
       \comma & a_0 = \cosh \rho \period
   \end{array}
\eqe

%
\par\bigskip\ni
{\it Bosonic fluctuation}
\par\medskip

In the conformal gauge, the bosonic part of the world-sheet action
is given by \eqb
   S_B &=& -\frac{\sqrt{\lambda}}{4\pi} \int d\tau d\sigma \, \eta^{ij}
     G_{\mu\nu} \del_i x^\mu \del_j x^\nu
    \comma \label{SB}
\eqe where $i,j = (\tau,\sigma)$, and  $\eta_{\tau\tau} = -
\eta_{\sigma\sigma} = +1$. A simple way to obtain the fluctuation
Lagrangian is the geodesic expansion \cite{NC}. Introducing a
vector $y^{a}$ in the tangent space of the space-time, the
quadratic fluctuation Lagrangian is given by \eqb
   L_B^{(2)} &=& - \half \eta_{ab} D_i y^a D^i y^b
     - \half y^a y^b e_i^c e^{i \, d} R_{acbd} \label{LB2}
   \period
\eqe $\eta_{ab}$ is the flat ten-dimensional metric with mostly
minus signatures. $D_{i}$ and $e_{i}^{a}$ are the projections of
the covariant derivative $D_{\mu}$ and the vielbein $e^{a}_{\mu}$,
respectively; for example, $(D_{i} y)^{a} = \del_{i}
x^{\mu}(\del_\mu + \omega_{\mu, b}^{\ \ \  a}) y^{b}$ with
$\omega_{\mu,ba}$ the connection one-form. For $AdS_{5} \times
S^{5}$, the curvature $R_{abcd}$ is simple: \eqb
   R_{abcd} = \mp (\eta_{ac}\eta_{bd} - \eta_{ad}\eta_{bc})
     \comma \label{Reta}
\eqe with the minus sign if all  indices $(a,b,c,d)$ correspond to
$AdS_{5}$ and the plus sign if $(a,b,c,d)$ to $S^{5}$; otherwise
it vanishes. In term of the global coordinate system in
(\ref{AdS5S5}), the non-vanishing $\omega_{i,ab} (= \del_{i}
x^{\mu} \omega_{\mu,ab})$ are, up to the anti-symmetry, \eqb
   \begin{array}{lll}
    \omega_{i,\theta\rho} = \cosh \rho \del_i \theta \comma
  & \omega_{i,t \rho} = -\sinh \rho \del_i t \comma
  & \omega_{i,\phi_4\rho} = \cosh \rho \cos \theta \del_i \phi_4 \comma \\
  \omega_{i,\phi_4\theta} = -\sin \theta \del_i \phi_4 \comma
 & \omega_{i,\phi_5\rho} = \cosh \rho \sin \theta \del_i \phi_5 \comma
  &\omega_{i,\phi_5\theta} = \cos \theta \del_i \phi_5 \comma \\
  \omega_{i,\psi\gamma} = \cos \gamma \del_i \psi \comma
  & \omega_{i,\phi_3\gamma} = -\sin \gamma \del_i \phi_3 \comma
  & \omega_{i,\phi_1\gamma} = \cos \gamma \cos \psi \del_i \phi_1 \comma \\
    \omega_{i,\phi_1\psi} = -\sin \psi \del_i \phi_1 \comma
 & \omega_{i,\phi_2\gamma} = \cos \gamma \sin \psi \del_i \phi_2 \comma
  &\omega_{i,\phi_2\psi} = \cos \psi \del_i \phi_2 \period
 \end{array}
\eqe

\par\bigskip\ni
{\it Fermionic fluctuation}
\par\medskip

To study the fermionic fluctuations, we use the quadratic
fermionic part of the type IIB Green-Schwarz (GS) action on $AdS_5
\times S^5$ \cite{MT}. Following the notation in \cite{HW}, it is
expressed as \eqb
    L^{(2)}_F &=& i \theta^I P_{IJ}^{ij} \calD_{ij JK} \theta^K
  \comma  \\
  && \calD_{ij JK} = \sighat_i \lbb
     (\del_j - \frac{1}{4} \omega_{j,ab} \sigma^{ab}) \delta_{JK}
   - \frac{1}{4\cdot 480} {F}_{abcde} \sigma^{abcde} \sighat_j
   (\rho_0)_{JK}
   \rbb \comma \nn \\
 && P_{IJ}^{ij}  = \eta^{ij}\delta_{IJ} + \epsilon^{ij}(\rho_3)_{IJ} \comma
    \quad \sighat_j = e_j^a \sigbar_a \comma \quad \epsilon^{01} = +1
   \comma \nn \\
 &&  \rho_0 = \matrixii{0}{1}{-1}{0}
    \comma \quad \rho_3 = \matrixii{1}{0}{0}{-1} \period \nn
\eqe Here, $(\theta^I)^{\alpha} \ (I = 1,2)$ are ten-dimensional
Majorana-Weyl spinors with 16 components;
$(\sigbar^a)_{\alpha\beta}, (\sigma^a)^{\alpha\beta}$ are $16
\times 16$ gamma matrices in ten dimensions; their
anti-symmetrization is given, e.g.,  by
 $\sigma^{ab} = (\sigma^{a}\sigbar^{b} - \sigma^{b}\sigbar^{a})/2$,
$ \sigbar^{ab} = (\sigbar^{a}\sigma^{b} -
\sigbar^{b}\sigma^{a})/2$. If we label the $AdS_5$ and $S^5$ parts
by $a,b = (0,6,7,8,9)$ and $(1,2,3,4,5)$, respectively, the
non-vanishing components of the five-form are $ {F}_{06789} =
{F}_{12345} = 4$.

The GS action has the $\kappa$-symmetry, the relevant part of
which is now \eqb
  \delta_\kappa \vartheta^\alpha
   = (\sighat_i)^{\alpha \beta} \eta^{ij}
  \kappa_{j \beta} \comma
\eqe where $\eta^{ij} \kappa_{j \beta} = \epsilon^{ij}
\bar{\kappa}_{j \beta}$,
 $\vartheta = \theta^1 + i \theta^2 $ and
$\kappa_{j \beta} = \kappa^1_{j \beta} + i \kappa^2_{j \beta}$.
Following \cite{FT2}, we fix this symmetry by setting \eqb
   \theta^{1} = \theta^{2} \period
\eqe One can check that this gauge is actually possible for the
backgrounds which we consider in the following sections. In this
gauge, the form of the quadratic Lagrangian is simplified to \eqb
   L^{(2)}_F &=& 2i \theta^1 D_F \theta^1 \comma \nn \\
   D_F &=& \eta^{ij}\sighat_i  (\del_j - \frac{1}{4} \omega_{j,ab} \sigma^{ab})
    - \frac{1}{2}\epsilon^{ij} \sighat_i \sigma_* \sighat_j \comma \label{DF}
\eqe where $\sigma_{*} \equiv \sigma^{06789} = \sigma^{12345}$.

\msection{Three-spin rotating string in $S^5$}

In this paper, we consider classes of the ``constant radii"
solutions \cite{Tseytlin}, in which $a_r$ in (\ref{Zr}) are
constant. Bearing in mind possible extensions to more general
cases, we develop a method to compute one-loop effective actions
in a large-spin expansion. In this and the next sections, we
discuss it in some detail for the constant radii strings with
three spins in $S^5$, i.e., in the $SO(6)$ sector. We apply this
method to other cases later.

\par\bigskip
\msubsection{Solution}

The solution which we consider is \eqb
    Z_0 = e^{i\kappa \tau} \comma \quad Z_s = a_s\, e^{i(w_s \tau + m_s \sigma)}
   \comma \label{3spin1}
\eqe where $s=1,2,3$; $\kappa, a_s, w_s$ are constant with
$\Sigma_s a_s^2 = 1$; $m_s$ are integers (when the period of
$\sigma$ is $2\pi$). Other fields including fermionic ones vanish.
The equations of motion and the Virasoro constraints give \eqb
    && w_s^2 = \nu^2 + m_s^2  \quad ({\rm if } \ a_s \neq 0) \comma \nn \\
    && \kappa^2 = \sum a_s^2 ( w_s^2 + m_s^2) \comma \quad
  0 = \sum a_s^2 w_s m_s  \comma \label{3spin2}
\eqe with $\nu$ a constant.  Classically, the space-time energy
and the three spins in $S^{5}$ are given by \eqb
   E = \sql \kappa \comma \quad J_{s} = \sql a_{s}^{2} w_{s} \comma
    \label{clcharge}
\eqe where $\sql = R^{2}/\alpha'$ and $R$ is the radius of
$AdS_{5} \times S^{5}$.

When $J_{1} = J_{2}$ and $m_{3} = 0$, the parameters of the
solution become \eqb
  && w_1 = w_2 \equiv w \comma \quad m_1 = - m_2 \equiv m \comma \quad
   a_1 = a_2 = \sg/\sqrt{2}
  \comma \nn \\
   && w_3 = \nu \comma \quad m_3 = 0 \comma \quad a_{3} = \cg \comma
     \label{J1eqJ2} \\
   && w^2 = \nu^2 + m^2 \comma \qquad \kappa^2 = \nu^2 + 2 m^2 \sg^2
    \comma \nn
\eqe where $ s_{x} \equiv \sin x , c_{x} \equiv \cos x$. In the
following, we call this simplified solution the $J_{1}= J_{2}$
three-spin solution. The point-like (BMN) solution in \cite{BMN}
is also obtained by setting $ a_{1,2} =w_{1,2} = m_{1,2} = m_{3}
=0$ in (\ref{3spin1}) and (\ref{3spin2}).

\par\bigskip
\msubsection{Bosonic fluctuation}

Now, let us consider the bosonic fluctuations around the
three-spin solution in (\ref{3spin1}). Substituting the solution
into (\ref{LB2}), one finds that the fluctuations in the $AdS_{5}$
and the $S^{5}$ parts decouple. The contribution to the one-loop
effective action from the $AdS_{5}$ part is then represented by
the determinant of the quadratic operator, \eqb
   D^B_{pq} &=& -\eta_{pq} \del^2 + \kappa^2 R_{ptqt}  \comma \label{Dpq}
\eqe where $p,q = (t,\rho, \theta, \phi_{4},\phi_{5}) $ or $
(0,6,7,8,9)$, and $\del^{2} = \eta^{ij}\del_{i}\del_{j}$. This
operator describes one time-like massless boson and four
space-like bosons with mass squared $\kappa^{2}$, as in the BMN
case. Taking the determinant with respect to the tangent space
indices gives \eqb
   \det D^B_{pq} = -\del^{2}(\del^{2} + \kappa^{2})^{4} \period \label{detDpq}
\eqe

For the $S^{5}$ part,  the quadratic term of the connection
one-form and the curvature term cancel each other, to give \eqb
  D_{mn}^B &=&  \delta_{mn} \del^2
    + 2(\omega_{\tau,mn} \del_\tau - \omega_{\sigma,mn} \del_\sigma )
      \comma \label{DmnB} \\
   \det D_{mn}^B & = &
      \del^2 \Bigl[ (\del^2)^4 +
   \Bigl( \Sigma_{s=1}^3 (1-a^2_s) \Omega_s^2 \Bigr) (\del^2)^2
        +  (a_1^2 \Omega_2^2 \Omega_3^2 + a_2^2 \Omega_3^2 \Omega_1^2
        + a_3^2 \Omega_1^2 \Omega_2^2 )
   \Bigr]  \comma \nn
\eqe where $m,n = (\phi_{1},\phi_{2},\phi_{3},\psi,\gamma)$ or
$(1,2,3,4,5)$, and \eqb
   \Omega_s = 2 (w_s\del_\tau - m_s \del_\sigma) \period
\eqe This determinant has been obtained in \cite{ART}. Note that
we have assumed here that none of $a_{s}$ vanishes in order to use
the constraints $w_{s}^{2} = \nu^{2} + m_{s}^{2}$.

\par\bigskip\ni
\ni {\it Change of functional bases}
\par\medskip

Later, we explicitly evaluate the functional determinant in a
large $ J(=\sum J_{s})$ expansion. It turns out that $\det
D_{mn}^{B}$ in (\ref{DmnB}) has an inappropriate infrared behavior
for this purpose. Here, we make an $SO(5)$ rotation to avoid it,
concentrating on the $J_{1}=J_{2}$ three-spin solution with
(\ref{J1eqJ2}).

First, we take an orthogonal  matrix, \eqb
    Q^{m}_{\  n}
        = \frac{1}{\sqrt{2}v}\lb \begin{array}{ccccc}
          \sg \nu  &  v & 0 &  \cg w & 0 \\
          \sg \nu   &  -v & 0 & \cg w & 0 \\
          \sqrt{2} \cg w  & 0 & 0  &  -\sqrt{2}\sg \nu & 0 \\
          0  &  0 & \sqrt{2}v &  0 & 0 \\
          0  & 0 &  0 & 0 & -\sqrt{2} v
        \end{array}
        \rb \comma
\eqe with $v^{2} = w^{2} - M^{2}$ and $M^{2} = \sg^{2} m^{2}$. By
a change of bases using $Q^{m}_{\  n}$, the connection one-form is
brought into a standard form. Namely, defining $ \omhat_{i,mn}
\equiv Q^{k}_{\ m} \omega_{i,kl} Q^{l}_{\ n}$, we find that \eqb
   \omhat_{\tau,mn}
     = \lb
       \begin{array}{ccc}
           0 & &  \\
             & p(-w) &  \\
             & & p(-v)
       \end{array}
    \rb  \comma \qquad
    p(x) = \matrixii{0}{x}{-x}{0}
         \period
\eqe We further introduce \eqb
   R^{m}_{\ n}(\tau) =
    \lb  \begin{array}{ccc}
            1  &   &  \\
               &  P(\tau w) & \\
               & & P(\tau v)
     \end{array}
     \rb \comma \qquad
     P(x) = \matrixii{\cos x}{\sin  x}{-\sin  x}{\cos x}
      \comma \label{RP}
\eqe so that $ R^{m}_{\ n} $ satisfies \eqb
   0 = \del_\tau
   R_{mn} -  \omhat_{\tau,mk} R^{k}_{\ n}
   \period
\eqe Then, by an transformation of the form, \eqb
   \hat{\calO} \equiv (QR)^{-1} \calO (QR)
     \comma \label{hatQR}
\eqe the quadratic operator becomes \eqb
   \Dhat_{mn}^{B} & = & R^{k'}_{\ m} Q^{k}_{\ k'} D^B_{kl} \, Q^l_{\ l'} R^{l'}_n  \nn \\
         &=&   \delta_{mn} \del^2  + M_{mn}
            - 2 \rho_{\sigma,mn} \del_\sigma \comma \label{DhatM}
\eqe where $M_{mn} = {\rm diag} (0,w^2,w^2,v^2,v^2)$ and $
\rho_{\sigma,mn} = R^{k}_{\ m} \omhat_{\sigma, kl} R^{l}_{\ n}$.
After some algebra, we also find that \eqb
    \det \Dhat_{mn}^{B}
   &=& (\del^2 + w^2)^2 (\del^2 + v^2)^2\del^{2} \nn \\
    && \ + \, 4 \cg^2 k^2 \del_\sigma^2 \del^{2}
    \Bigl[ (1+\frac{w^2}{v^2})(\del^2 + w^2)(\del^2 + v^2)
    + 4 \cg^2k^2 (\frac{w}{v})^2 \del_\sigma^2 \Bigr]
   \\
  && \  + \,  4 \sg^2 k^2 (\frac{\nu}{v})^2 \del_\sigma^2  (\del^2 + v^2)
   \Bigl[ (\del^2+w^2)(\del^2+v^2)
     + 4 \cg^2 k^2 \del_\sigma^2 \Bigr] \nn \period
\eqe

\msubsection{Fermionic fluctuation}

Now, let us move onto the fermionic part. In order to evaluate the
one-loop determinant, it is useful to make a rotation of $ D_F $,
so that the kinetic term is simplified to take the form for
two-dimensional fermions \cite{LT}. For this purpose, we introduce
an element of $SO(1,9)$, \eqb
  Q^{a}_{\ b} = \matrixii{q^a_{\ b}}{0}{0}{1_{6\times 6}} \comma \quad
  q^a_{\ b} = \frac{1}{M} \lb
    \begin{array}{cc}
       \kappa & - W_s  \\
      0 & M_s \\
      -W & \kappa W_s/W \\
       0 & l_s M
    \end{array}
  \rb \comma \label{Q1}
\eqe where $a,b = (t, \phi_1, \phi_2, \phi_3, ...)$; $W_{s} \equiv
a_{s} w_{s}, M_{s} = a_{s} m_{s}$, $W^{2} \equiv \sum_{s}
W_{s}^{2}, M^{2} \equiv \sum_{s} M_{s}^{2}$; $1_{6\times6}$ is the
$6 \times 6$ unit matrix; $  l_{s_1} = \ep_{s_1s_2s_3}
M_{s_2}W_{s_3}/MW$. This was chosen so that $ Q_{\tili}^{\ a} =
e^{-\varphi} e_{\tili}^{a}$ for $ \tili = (0,1)$, where $ \varphi$
is defined by the proportionality coefficient between the induced
metric, $h_{ij} = e_{i}^{a}e_{j}^{b} \eta_{ab}$, and the
world-sheet metric through $ h_{ij} = e^{2\varphi} \eta_{ij}$. In
our case, $e^{2\varphi} = M^{2}$. We then transform $D_F$ by the
element of $SO(1,9)$ corresponding to $ Q^{a}_{\ b}$, which we
denote by $S(Q)$. Since  $S^{-1} \sigma_{a} S = \sigma_{b} Q^b_{\
a}$ and hence \eqb
   S^{-1} \sighat_i S = e^\varphi \delta^{\tili}_i \sigbar_{\tili}
   \comma
\eqe this transformation replaces $\sighat_{i}$ with $\sigbar_{i}$
in the quadratic operator, to give a desired form. In the
following, we do not distinguish the indices $\tili$ and $i$.
After some algebra, we then find that
 \eqb
     S^{-1}(Q) D_F S(Q) =  e^{\varphi}\Dhat_F \comma
\eqe
 where
\eqb
  \Dhat_{F} &= & \eta^{ij}\sigbar_i  \Bigl[ \del_j - \frac{1}{4} \omega_{j,ab}
    (\sigma^{ab})' \Bigr]
    - \frac{1}{2}\epsilon^{ij} e^{\varphi}\sigbar_i \sigma'_* \sigbar_j \comma \nn \\
   & = &\sigbar^{i} \del_{i} + W \sigbar^{345} + \frac{1}{2MW}
      \sum_{a=0,1} \sum_{b=2,3} \sum_{c=4,5}
       \alpha_{abc} \, \sigbar^{abc}
        \comma \label{DF'}
\eqe $ (\sigma^{ab})' = S^{-1} \sigma^{ab} S, \sigma'_* = S^{-1}
\sigma_* S$, and \eqb
  &&  \alpha_{024}
    = - \sg \cg \kappa (\cp^2 m_1^2 + \sp^2 m_2^2  -m_3^2)
   \comma \qquad
     \alpha_{124} = (\cg/\sg) {\kappa} m_3 w_3 \comma \nn \\
  &&  \alpha_{034}   = \sg \sp \cp w_3 (m_1 w_2 - m_2 w_1)
   \comma \qquad \ \ \ \
    \alpha_{134}
            = {\sg \sp \cp }  {m_3} (m_1 w_2 - m_2 w_1) \comma  \nn \\
   && \alpha_{025}
             = {\sg \sp \cp} {\kappa } (m_1^2 -m_2^2)
    \comma \qquad \qquad \qquad \,
   \alpha_{125}
              = {\sg \sp \cp} {\kappa } (m_1 w_1 -m_2 w_2)
     \comma  \nn \\
    && \alpha_{035}
             =  - {\sg \cg}
       [ m_3 w_1 w_2 - w_3 (\cp^2 m_1 w_2 + \sp^2 m_2 w_1) ]
    \comma   \nn \\
&& \alpha_{135}
              =  {\sg \cg}
       [ m_1 m_2 w_3 - m_3 (\sp^2 m_1 w_2 + \cp^2 m_2 w_1) ]
   \period
\eqe In the above, we have assumed that none of $a_{s}$ vanishes.

Since $e^{\varphi}$ is constant, we have now only to evaluate the
determinant (pfaffian) of $\Dhat_{F}$. To proceed, we adopt the
following explicit realization of the gamma matrices (just to
evaluate the determinants with respect to the spinor indices):
\eqb
   \begin{array}{lll}
    \sigma^1 = \tau_3 \otimes 1 \otimes 1 \otimes 1 \comma &
    \sigma^4 = \tau_2 \otimes \tau_1 \otimes 1 \otimes 1 \comma &
    \sigma^7 = \tau_2 \otimes \tau_2 \otimes \tau_2 \otimes \tau_3 \comma \\
    \sigma^2 = \tau_1 \otimes 1 \otimes 1 \otimes 1 \comma  &
    \sigma^5 = \tau_2 \otimes \tau_2 \otimes \tau_3 \otimes 1 \comma &
    \sigma^8 = \tau_2 \otimes \tau_2 \otimes \tau_2 \otimes \tau_1 \comma  \\
    \sigma^3 = \tau_2 \otimes \tau_3 \otimes 1 \otimes 1 \comma &
    \sigma^6 = \tau_2 \otimes \tau_2 \otimes \tau_1 \otimes 1   \comma &
    \sigma^9 = \tau_2 \otimes \tau_2 \otimes \tau_2 \otimes \tau_2  \comma
      \end{array}
  \label{sigmas}
\eqe $\sigma^{0}=\sigbar^{0} = 1$, and $\sigbar^{a} = -
\sigma^{a}$ $(a=1, ..., 9) $, where $\tau_{a}$ are the Pauli
matrices. With these gamma matrices, $\Dhat_{F}$ takes the form
\eqb
    \Dhat_F & = & \matrixii{\Delta_{F}^{+}}{0}{0}{\Delta_{F}^{-}}
     \otimes 1 \period
\eqe Denoting by the same symbols the matrices which are obtained
by extracting the first two matrices in the tensor products in
(\ref{sigmas}), for example, $\sigma^{1} \to \tau_{3} \otimes 1$,
one finds that \eqb
   \Delta_{F}^{\pm} &=& \sigbar^{i} \del_{i}
     \mp W \sigbar^{012}
      + \beta_{1\pm} \sigbar^{024}
       +   \beta_{2\pm} \sigbar^{124}
       + \beta_{3\pm} \sigbar^{034}
       + \beta_{4\pm} \sigbar^{134} \comma \label{DeltaF}
\eqe where \eqb
  &&    \beta_{1\pm} = \frac{1}{2MW} (\alpha_{024} \pm \alpha_{135}) \comma \quad
          \beta_{2\pm} =  \frac{1}{2MW} (\alpha_{124} \pm \alpha_{035}) \comma \nn  \\
   &&    \beta_{3\pm} = \frac{1}{2MW} (\alpha_{034} \mp \alpha_{125}) \comma \quad
          \beta_{4\pm} =  \frac{1}{2MW} (\alpha_{134} \mp \alpha_{025}) \period
\eqe From this, it follows that \eqb
  \det  \Delta_F^\pm & \! \! = \! \!&
      (\del^{2})^2 + 2 W^2 \del^{2}
    + 2 ( \sum_{n=1}^{4}\beta_{n\pm}^{2} ) (\del_\tau^2 + \del_\sigma^2)
     + 4(\beta_{1\pm}\beta_{2\pm} +
     \beta_{3\pm}\beta_{4\pm})\del_\tau\del_\sigma
      \label{detDeltaF} \\
  &&
    + \ (\sum_{n=1}^{4}\beta_{n\pm}^{2} )^{2}
     - 4 (\beta_{1\pm}\beta_{2\pm} + \beta_{3\pm}\beta_{4\pm})^{2}
     + 2 (\beta_{1\pm}^{2} - \beta_{2\pm}^{2} - \beta_{3\pm}^{2} + \beta_{4\pm}^{2}) W^{2}
     + W^{4} \period \nn
\eqe The final one-loop contribution of the fermionic sector is
then represented by \eqb
    {\rm pf} \, \Dhat_F = \det\Delta_F^+ \det\Delta_F^-
    \period \label{pfaffian}
\eqe Given this result, it would be interesting to generalize the
analysis in \cite{FPT,BTZ} to the generic three-spin constant
radii solution.

\par\bigskip\ni
{\it Change of functional bases}
\par\medskip

As in the case of $D_{mn}^{B}$, it turns out that the form of $
\det \Delta_{F}^{-}$ in (\ref{detDeltaF}) is not convenient for
our purpose. Thus, we make a change of bases again.
To this end, we introduce a $4 \times 4$ matrix \eqb
    R(\alpha) = 1 \otimes P(\tau \alpha) \comma \label{R1P}
\eqe with $P(x)$ given in (\ref{RP}). One then finds that $\det
R^{-1}(\alpha) \Delta_{F}^{\pm} R(\alpha)$ is given just by
replacing $\beta_{3\pm}$ in (\ref{detDeltaF}) with $ \beta_{3\pm}
- \alpha$. With this in mind, we define \eqb
   \Delhat_{F}^{\pm} \equiv R^{-1}(\beta_{3\pm}) \Delta_{F}^{\pm} R(\beta_{3\pm}) \comma
   \label{DelhatF}
\eqe so that $\beta_{3\pm}$ in (\ref{detDeltaF}) are set to be
zero. This transformation for $\Delta_{F}^{+}$ is not inevitable
for our purpose, but simplifies later computations.

Now, we focus on the $J_{1}=J_{2}$ three-spin solution. In this
case, \eqb
    \beta_{1\pm} = - \frac{\cg m (\kappa \pm \nu)}{2W} \comma \quad
    \beta_{3\pm} = \frac{w (\nu \mp \kappa)}{2W} \comma \quad
    \beta_{2\pm} = \beta_{4\pm} = 0 \comma \label{J1J2beta}
\eqe and hence \eqb
   \det \Delhat_{F}^{\pm} =
    (\del^2)^{2}+ 2 W^2 \del^2 + 2 \beta_{1\pm}^2 (\del_{\tau}^{2} + \del_{\sigma}^{2})
   + (\beta_{1\pm}^{2} + W^{2})^{2} \period
\eqe One can confirm that the original determinants in
(\ref{detDeltaF}) with (\ref{J1J2beta}) reproduce the
characteristic frequencies obtained in \cite{FT2}.\footnote{
Precisely, the world-sheet momenta here are integer moded, whereas
those in \cite{FT2} are half-integer moded: We have started with
an $su(2)$ rotated coordinates \cite{ART}, and do not need a
$\sigma$-dependent rotation.}

\msection{One-loop effective action and large $J$ expansion}
In this section, based on the results in section 3, we consider
the one-loop effective action of the GS string in the $J_{1}
=J_{2}$ three-spin  background. For this background, the
corresponding gauge theory operators have been identified
\cite{EMZ}, and there exit some parameter regions where the
fluctuations are stable \cite{ART,FT2}. We develop a large $J$
(total spin) expansion of the one-loop effective action, and
compute it in a closed form up to and including $\calO(1/J)$.

Collecting the contributions  from the bosonic, the fermionic and
the ghost sectors,  the one-loop effective action is given by \eqb
   e^{i \Gamma^{(1)}} =
    {\Det(\Delhat^+_F) \Det(\Delhat^-_F) \Det (-\del^{2})
  \o \Det^{\half}(D_{pq}^{B}) \Det^{\half}(\Dhat_{mn}^{B}) }
  \comma \label{S1}
\eqe where $\Det$ stands for the functional determinant. To
develop a large $J$ expansion, we expand the fluctuation operators
with respect to $\nu$ with $\del_{i}$ and $m$ fixed: \eqb
   \del^{2} + \kappa^{2} &=& (\del^{2} + \nu^{2}) + 2 M^{2}\comma \nn \\
   \det \Dhat_{mn}^{B} / \del^{2} &=&
     (\del^{2} + \nu^{2})^{4}
       + 2 \Bigl[ (2-\sg^{2}) + 2 \sg^{2} \frac{\del_{\sigma}^{2}}{\del^{2}} \Bigr]
     m^{2}(\del^2 + \nu^2)^3
      \label{expandD} \\
    && \qquad  + \
        8 \cg^2 m^2  \del_\sigma^2 (\del^2 + \nu^2)^2 + \cdots
       \comma \nn  \\
    \det \Delhat_{F}^{\pm}   &=&  (\del^{2} +\nu^{2})^{2} + 2 M^{2} \del^{2}
       + 2 \beta_{1\pm}^{2} (\del_{\tau}^{2} + \del_{\sigma}^{2}) + 2 \nu^{2} (M^{2} + \beta_{1\pm}^{2})
         + \cdots \period \nn
\eqe Note that $J = \Sigma_s J_s \sim \sql \nu$ and $M^2 \sim 2m^2
J_1/J$ for large $\nu$. This expansion is also regarded as that
with respect to the power of $ \del^2 + \nu^2$, or the power of
winding number $m$. The validity of the expansion becomes clear
shortly. Denoting the subleading terms for $ \det
\Dhat_{mn}^{B}/\del^2$ and $ \det \Delhat_{F}^{\pm}$ by $\delta
\Dhat_{mn}^{B}$ and $ \delta \Delhat_{F}^{\pm}$, respectively, and
plugging (\ref{expandD}) into (\ref{S1}), we find that the leading
terms cancel each other. This is a consequence of the asymptotic
BPS condition \cite{MMT}, or confirms the result for the
point-like BMN solution. We are then left with \eqb
   i \Gamma^{(1)} &=&
   \sum_{\eta = \pm }\Tr \log
   \Bigl( 1+ \frac{\delta \Delhat_F^\eta}{[\del^2 + \nu^2]^2}\Bigr)
      \label{S1Trlog} \\
  &&  \qquad - \ 2 \Tr \log
   \Bigl( 1+ \frac{2M^{2}}{\del^2 + \nu^2} \Bigr)
  - \half \Tr \log
 \Bigl(1+ \frac{\delta \Dhat^B_{mn}}{[\del^2 + \nu^2]^4}\Bigr)
   \period \nn
\eqe

We would like to evaluate these terms by expanding the logarithms
around unity. To check if that is possible,  we first consider
functional traces of the form \eqb
   \Tr \frac{(\del^2_\tau)^a (\del^2_\sigma)^b}{(\del^2)^c(\del^2+\nu^2)^d}
   \period \label{Trdd}
\eqe Since we are working with a Lorentzian world-sheet, we define
the trace by the standard $ i \ep$ prescription. We denote the
volumes of the world-sheet time and space by $T$ and $2\pi L$, the
derivatives $\del_{\tau} $ and $ \del_{\sigma}$ in the momentum
space by $ i\omega $ and $ i p_{n} = i n/L \ (n \! \in \! \bfZ)$,
respectively. Then,  the trace reads $ iT \sum_{n} \int {d\omega
\o 2\pi i}$, and the $\omega$-integral here picks up the poles on
the negative real axis. If one can approximate the summation of
$p_{n}$ by an integral, one easily finds the leading large $\nu$
behavior of the trace in (\ref{Trdd}) to be
\eqb
    TL { 1 \o \nu^{2(c+d-a-b-1)} }  \period \label{scaling}
\eqe

Here, we have to be a little careful about the infrared behavior:
The approximation by an integral requires that the derivatives of
the resultant integrand do not grow too large, but this may not
hold for $ p_n \sim 0$ because of the operator $1/\del^{2}$.
Fortunately, in our case, this ``massless'' operator appears
always in the combination $\del_{\sigma}^{2}/\del^{2}$. From this,
one can confirm, by evaluating the most singular terms for $p_{n}
\sim 0$, that the infrared behavior does not spoil the scaling in
(\ref{scaling}) for the terms which we consider. Using this
scaling, we find that the expansion of the logarithms in
(\ref{S1Trlog}) around unity actually gives a large $\nu $ (or
$J$) expansion.

More precisely, to evaluate the infrared behavior, we first divide
the summation over $n$ into two parts, i.e., that over $ {|n|
\siml \nu L}$ and $ {|n| \simg \nu L}$. The latter can be safely
approximated by an integral by scaling $\omega, p_n$ to
$\omega/\nu, p_n/\nu$.  The contribution which can be most
singular for $p_n \sim 0$ comes from the residue of the pole at
$\omega = -p_n$, and behaves as $ \sum_n \nu^{-2d}
p_n^{1+2(a+b-c)}$. Thus, for $ 2(c-a-b) -1 \leq 0$, this is
infrared finite and the approximation by an integral is allowed.

The validity of the expansion is confirmed also as follows: The $i
\ep$ prescription implies that the functional trace is given by a
continuation from the Euclidean case. Then, one can easily check
that, for large $\nu$, the ``subleading" terms in (\ref{expandD})
are indeed smaller than the leading terms (some powers of $(\del^2
+ \nu^2)$) irrespectively of the values of $\del^2_{\tau,\sigma}$.

We remark that if we use $ D_{mn}^{B}$ and $\Delta_{F}^{-}$
instead of $\Dhat_{mn}^{B}$ and $\Delhat_{F}^{-}$, we encounter
the operator $ (\del^{2})^{2} + 4\nu^{2} \del_{\tau}^{2}$ instead
of $ (\del^{2} + \nu^{2})^{2}$ for the leading terms, and its
infrared behavior prevent the expansion of the logarithms for the
large $J$ expansion. This is essentially the same problem as in
\cite{FT2} (see \cite{FPT}). In other words, for small
$\del_\tau$, $ (\del^{2})^{2} + 4\nu^{2} \del_{\tau}^{2}$ can be
smaller than other ``subleading" terms, which invalidates the
expansion.

Also, in general, one has to be careful about conjugation of a
differential operator $D$ by
a time-dependent operator in evaluating the determinant of $D$. However,
the time-dependent operators in our case are harmless, since they are
rotation operators (e.g., (\ref{RP})) which are essentially the same
as operators such as $e^{i c \tau}$ with $c$ a constant.
To see this, first note that such operators just shift a pair of characteristic
frequencies by a constant:
$\omega_n \to \omega_n \pm c$, and these shifts are irrelevant to
$ \log \Det \, D = \Tr \log D \sim \sum \omega_n$; the plus and minus
shifts cancel each other.
One can confirm this by a simple example. For example, by the conjugation using $P(\tau \nu)$,
$
  P^{-1}(\tau \nu) \matrixii{\del^2}{2\nu \del_\tau}{-2\nu
\del_\tau}{\del^2} P(\tau \nu) = {\rm diag} \, (\del^2 + \nu^2,
\del^2 + \nu^2)
$
and hence
\eqb
   \Det (\del^4 + 4 \nu^2 \del^2_\tau) = \Det(\del^2 + \nu^2)^2 \period \label{d4d2}
\eqe
The characteristic frequencies of $(\del^4 + 4 \nu^2 \del^2_\tau)$ are
$\omega_n = \sqrt{p_n^2 + \nu^2} \pm \nu$, whereas those of $(\del^2 + \nu^2)^2$
are $\omega_n = \sqrt{p_n^2 + \nu^2}$ with double degeneracy.
Thus, the sum of the characteristic frequencies is actually the same.
Second, note that the rotation operators may not change the norm
of the functions and hence the Hilbert space.
Finally, in the actual calculation, we first put a cut-off for large $|\tau|$
following the standard procedure of field theory. Then, the volume of the time
$T$ is factored out as in (\ref{S1M2}).
Therefore, subtleties about $\tau \to \pm \infty$, if any,
may not be relevant to the final results.

We are now ready to evaluate $\Gamma^{(1)}$. Once the expansion of
the logarithms is allowed, the problem reduces to an ordinary
calculation of an effective action in field theory.
In fact, our expansion is essentially the same as the large
${\rm (mass)}^2 \sim \nu^2$ expansion which guarantees the decoupling
of heavy particles. This enables
us to systematically compute $\Gamma^{(1)}$ up to higher order
terms in the $1/\nu$ expansion.
One then finds that, up to and
including $\calO(1/\nu)$,
 the first expansion of the logarithms is enough, and
the terms in the ellipses in (\ref{expandD}) do not contribute.
Denoting each term in (\ref{S1Trlog}) by $I_{F}^{\pm},
I_{AdS_{5}}$ and $I_{S^{5}}$, respectively,  they are
\eqb
   && I_{F}^{\pm} \sim
       iT \sum_{n \in \bfZ }\biggl[ \frac{M^2}{\sqrt{p_{n}^2 + \nu^2}}
        + \frac{\nu^{2}\beta_{1\pm}^{2}}{ (p_{n}^2 + \nu^2)^{ \frac{3}{2} } }
       \biggr] \comma \quad
   I_{AdS_{5}}  \sim
       iT\sum_{n \in \bfZ} \frac{-2M^2}{\sqrt{p_{n}^2+\nu^2}} \comma \label{3I} \\
    &&   I_{S^{5}}  \sim
      iT \sum_{n \in \bfZ} \biggl[  \frac{(\sg^{2}-2)m^{2}/2}{\sqrt{p_{n}^2+\nu^2}}
    +  \cg^{2}m^{2} \frac{ p_{n}^2}{(p_{n}^2+\nu^2)^{\frac{3}{2}}}
    - \frac{\sg^{2} m^{2}}{\nu^2} \biggl( \frac{p_{n}^2}{\sqrt{p_{n}^2+\nu^2}} - |p_{n}|
      \biggr)  \biggr]
  \comma \nn
\eqe
where $\beta^{2}_{1+} \sim \cg^{2} m^{2}, \beta^{2}_{1-} \sim
0$ at this order. Combining all, we arrive at
\eqb
  i \Gamma^{(1)} \sim  -iTL M^{2} \cdot C
   \comma \label{S1M2}
\eqe
with
\eqb
   C = \frac{1}{L\nu^{2}} \sum_{n \in \bfZ}
 \biggl( \frac{p_{n}^2+ (\nu^2/2)}{\sqrt{p_{n}^2+\nu^2}}  - |p_{n}| \biggr)
   \period \label{Cpn}
\eqe
One can check that $C$ is ultraviolet finite, in accord with
\cite{DGT}. It is also possible to evaluate $C$ in a closed form
for large $\nu$. Summing up  the summand as indicated above
(namely, first summing the terms in the parenthesis for given $n$
and then summing over $n$) can be approximated by integration, and
gives $C \sim 1/2$. We summarize the evaluation of $C$ in the
appendix.

Here, there may be some issues to be considered. One is about
regularization: Although $C$ is finite in the above combination,
each term in (\ref{S1Trlog}) or (\ref{3I}) is divergent. (Recall
that the final result is obtained by (infinite bosonic
contributions) $-$ (infinite fermionic contributions). )
 Thus, one needs to regularize them, and the value of $C$ may change with
a different regularization. For instance, if we adopt the
zeta-function regularization  for  each divergent sum in
(\ref{Cpn}), we obtain $C \sim -1/2$. (See the appendix.) Another
related issue is about finite renormalization: Since there are
infinities in the intermediate steps, one may also need to take
into account, in general, possible finite renormalization or
contributions from finite counter terms. In a
renormalizable theory, observables do not depend on regularization
and renormalization procedure. Ambiguities of finite parts are
fixed by appropriate renormalization conditions so as to maintain
the symmetry. We will return to these issues at the end of section 7.

\msection{$SU(2)$ sector}

In this section, we consider a class of solutions which is
obtained by setting in (\ref{3spin1}) and (\ref{3spin2}) \eqb
   a_{3} = w_{3} = m_{3} = 0 \label{twospin}
\eqe (with $J_{1} \neq J_{2}$ generally). This has two spins in
$S^5$, and is called the $SU(2)$ sector. The corresponding gauge
theory operators have been identified in \cite{KMMZ}. Since $w_{3}
= m_{3}=0$ but $ \nu \neq 0$ generally, one may not use relations
such as $w_{3}^{2} = \nu^{2} + m_{3}^{2}$, which have been used so
far (even before specializing to the $J_{1} = J_{2}$ three-spin
case). Thus, we repeat the same procedure from the beginning for
this much simpler case, and just display the results.

Let us begin with the bosonic sector.  The fluctuations in the
$AdS_{5}$ part are described by (\ref{Dpq}), and its determinant
by (\ref{detDpq}). For the $S^{5}$ part, the quadratic operator
takes the form \eqb
    D_{mn}^B =  \delta_{mn} \del^2
    + 2(\omega_{\tau,mn} \del_\tau - \omega_{\sigma,mn} \del_\sigma )
    + N_{mn} \comma
\eqe where the non-vanishing $\omega_{i,mn}$ are $\omega_{i,\phi_1
\psi} = -a_2 \del_i \phi_1 \comma
  \omega_{i,\phi_2 \psi} = a_1 \del_i \phi_2 $;
$N_{mn}$ is a diagonal matrix with $N_{\gamma\gamma} =
N_{\phi_{3}\phi_{3}} = \nu^{2}$ and others zero. This is different
from the corresponding operator in (\ref{DmnB}) by this mass term
(in addition to the form of $\omega_{i,mn}$).
 Its determinant is
\eqb
  \det D^B_{mn} = \del^2 (\del^2 + \nu^2)^2
   [\del^4 + a_1^2 \Omega_2^2 + a_2^2 \Omega_1^2] \period
\eqe Since  $D_{mn}^{B}$ is non-trivial only for $m,n =
(\phi_{1},\phi_{2},\psi)$, we first focus on this part. Similarly
to the previous case, we then make a transformation  of the type
(\ref{hatQR}) with \eqb
   Q^{m}_{\ n}
 = \frac{1}{\tilw}
  \matrixiii{a_{1}w_{2}}{a_{2}w_{1}}{0}{a_{2}w_{1}}{-a_{1}w_{2}}{0}{0}{0}{\tilw}
      \comma \quad
   R^{m}_{\ n} = \matrixii{1}{0}{0}{P(\tau\tilw)}   \comma
\eqe and $\tilw^{2} \equiv a_{1}^{2} w_{2}^{2} +
a_{2}^{2}w_{1}^{2} = \nu^{2} + a_{1}^{2} m_{2}^{2} +
a_{2}^{2}m_{1}^{2}$. This gives the same form of the quadratic
operator as in (\ref{DhatM}) with $M_{mn} = {\rm diag}
(0,\tilw^{2},\tilw^{2})$. Combining it with the trivial part from
$m,n= (\gamma, \phi_{3})$, we find that \eqb
   \det \Dhat_{mn}^{B}
     & = & \del^{2}(\del^{2} + \nu^{2})^{2} \Bigl[ (\del^{2} + \tilw^{2})^{2}
      + A_1{ \del_{\sigma}^{2} \o \del^{2} }
       (\del^{2} + \tilw^{2}) + A_2 \del_{\sigma}^{2}  \Bigr] \comma
\eqe where \eqb
   A_{1} =  \biggl[ \frac{2(m_{1}w_{1} + m_{2} w_{2})}{\tilw} \biggr]^2
     \comma \quad
   A_{2} = \biggl[ \frac{2a_{1}a_{2}(m_{1}w_{2} - m_{2} w_{1})}{\tilw} \biggr]^2
    \period
\eqe

Let us move on to the fermionic part. The quadratic fluctuations
are described by the operator (\ref{DF}). We then make an
$SO(1,9)$ transformation as before. In this case, the
corresponding element of $SO(1,9)$ turns out to be given by
(\ref{Q1}) with the two-spin condition (\ref{twospin}). (Note that
$l_{s}$ becomes $(0,0,1)$.) Proceeding similarly, one also finds
that the pfaffian of the transformed operator $\Dhat_{F}$ is given
by the same formulas,  (\ref{detDeltaF}) and (\ref{pfaffian}),
with \eqb
  &&  \beta_{1\pm} = \beta_{2\pm} = 0 \comma \\
  &&   \beta_{3\pm} =  \mp \frac{a_1a_2}{2MW}\kappa(m_1w_1-m_2w_2) \     \comma \quad
   \beta_{4\pm} =  \mp \frac{a_1a_2}{2MW} \kappa (m_1^2-m_2^2)
    \period \nn
\eqe ($M, W$ are defined as in three-spin case.) To simplify the
expression, we further make a rotation (\ref{DelhatF}), so that
$\beta_{3\pm}$ are removed. We then arrive at \eqb
   \det \Delhat_F^\pm
    =  (\del^{2} + \nu^{2})^{2} + 2M^{2} \del^{2} + 2 \beta_{4}^{2} (\del_{\tau}^{2}
    + \del_{\sigma}^{2})
      + (2\nu^{2} +\beta_{4}^{2} + M^{2})(\beta_{4}^{2} + M^{2}) \comma
\eqe where $\beta_{4}^{2} \equiv \beta_{4\pm}^{2}$.

Given these results, one can compute the one-loop effective action
for the two-spin case. Up to and including $\calO(1/\nu)$, it is
given by the same formula as  (\ref{S1M2}) and (\ref{Cpn}). In
this case, $M^{2} = a_{1}^{2} m_{1}^{2} + a_{2}^{2}m_{2}^{2} \sim
\alpha(1-\alpha)(m_1-m_2)^2$, where $\alpha \equiv J_1/(J_1+J_2)$.

The fluctuations of the $SU(2)$ case are not stable. However, we
did not see any signals of instability. We also note that the
results for the two-spin case with $J_{1} = J_{2}$ are obtained by
setting $\cg=0 \ (J_{3} = 0)$ in the $J_{1}=J_{2}$ three-spin
results. These suggest that our computation gives a smooth
continuation from the stable case, at least at this order.

\msection{$SL(2)$ sector}

There is another interesting class of simple solutions which has
one spin in $AdS_5$ and one in $S^5$ \cite{Tseytlin}. This is
called the $SL(2)$ sector. The fluctuations around these solutions
are stable for large spins. The corresponding gauge theory
operators have been identified in \cite{KZ}.

\par\bigskip
\msubsection{Solution}

The constant radii solution in the $SL(2)$ sector is given by \eqb
  Z_0 = a_0 \, e^{i \kappa \tau} \comma \quad Z_5 = a_5 \, e^{i(u\tau + k \sigma)}
    \comma \quad Z_1 =  a_1 e^{i(w\tau + m \sigma)} \comma
\eqe where $a_0,a_1,a_5,\kappa,u,w$ are constant with
 $a_0^2 - a_5^2 = 1$ and $a_1 = 1$; $k,m$ are integers (when
 the period of $\sigma$ is $2\pi$).
 Other fields are vanishing.
The equations of motion give a constraint \eqb
   u^2 = \kappa^2 + k^2 \period \label{eq1}
\eqe Introducing $(W_5, W_1) \equiv (a_5 u, w) $ and $ (M_5, M_1 )
\equiv (a_5 k, m)$, the Virasoro constraints read \eqb
   a_0^2 \kappa^2 = W^2 + M^2 \comma \qquad 0 = W \cdot M
     \period \label{sl2Vir}
\eqe The conserved charges are the space-time energy, $ E = \sql
a_0^2 \kappa$, and the spins in $AdS_5$ and $S^5$, $S = \sql a_5^2
u $ and $J = \sql w $, respectively.

In the following, we are interested in the large spin limit with
$k$ and $\alpha \equiv S/J$ fixed. Since the number of the
independent parameters are three, we are left with only one free
parameter. We take $\nu^2 \equiv w^2 - m^2$ as this free
parameter, and consider the large $\nu$ limit. Some useful
relations are $m = - \alpha k$ and, for large $\nu$, \eqb
  \begin{array}{ll}
   { \displaystyle
   \kappa \sim \nu[  1 + \frac{k^2}{\nu^2} \alpha(\alpha+1) ] }\comma
    & { \displaystyle u \sim \nu [ 1 + \frac{k^2}{2\nu^2} (2\alpha^2 + 2\alpha +1) ] } \comma \\
    { \displaystyle w \sim \nu[ 1 + \frac{k^2}{2\nu^2} \alpha^2 ]} \comma
  & { \displaystyle a_5^2 \sim \alpha [ 1 - \frac{k^2}{2\nu^2} (1+\alpha)^2 ]} \comma \\
  { \displaystyle M^2 \sim \alpha(1+\alpha) k^2 [ 1 - \frac{k^2}{2\nu^2} (1+\alpha) ]} \period &
    \end{array} \label{sl2largenu}
\eqe Note also that $ J \sim \sql \nu$.

\par\bigskip
\msubsection{Bosonic fluctuation}

From (\ref{LB2}), it is straightforward to read off the bosonic
fluctuation operator: The $S^5$ part is given by \eqb
    D_{mn}^B = {\rm diag} (\del^2, \del^2 + \nu^2, \del^2 + \nu^2, \del^2 + \nu^2,
     \del^2 + \nu^2) \comma
\eqe with $m,n = (\phi_1,\phi_2,\phi_3, \psi,\gamma) $, whereas
the $AdS_5$ part is \eqb
   D_{pq}^B &=& \matrixii{D_{pq}^1}{0}{0}{D_{pq}^2} \comma \label{DBpq} \\
   D_{pq}^1 &=&  - \eta_{pq} \del^2 + 2 \omega_{i,pq} \del^i \comma \quad
   D_{pq}^2 = {\rm diag} (\del^2 + \kappa^2, \del^2 + \kappa^2)
   \comma \nn
\eqe with $ p,q = (t,\phi_5,\rho,\theta, \phi_4)$. The
non-vanishing $\omega_{i,pq}$ are $\omega_{i,t\rho} = - a_5 \del_i
t $ and $ \omega_{i,\phi_5 \rho} =  a_0 \del_i \phi_5$. Thus, \eqb
   \det D_{mn}^B &=& \del^2(\del^2 + \nu^2)^4 \comma \\
   \det D_{pq}^B &=&  -\del^2(\del^2 + \kappa^2)^2[\del^4 -4a_5^2 \kappa^2 \del_\tau^2
         + 4 a_0^2 (u \del_\tau - k \del_\sigma)^2 ] \period \nn
\eqe These give the same characteristic frequencies as in
\cite{PTT}.

For the large $\nu$ (or $J$) expansion, we further make an
$SO(1,2)$ rotation of the type (\ref{hatQR}) for the $
(t,\phi_5,\rho) $ part. In this case, \eqb
   Q^p_{\ q} = \frac{1}{\tilu}
   \matrixiii{a_0 u}{a_5 \kappa}{0}{a_5 \kappa}{a_0 u}{0}{0}{0}{\tilu}
   \comma \quad
   R^p_{\ q} = \matrixii{1}{0}{0}{P(-\tilu\tau)} \comma
\eqe with $ \tilu^2 = a_0^2u^2 - a_5^2 \kappa^2$ and $P(x)$ given
in (\ref{RP}). Then, the fluctuation operator becomes \eqb
  \Dhat^1_{pq} = - \eta_{pq} \del^2
      + M_{pq} - 2 \rho_{\sigma,pq} \del_\sigma \comma \label{DhatBpq}
\eqe where $ M_{pq} =$ diag $(0,\tilu^2,\tilu^2)$ and
$\rho_{\sigma,pq} $ are defined similarly to the previous cases.
Combining this with the trivial part from $p,q=(\theta,\phi_4)$,
we find that \eqb
   \det \Dhat^B_{pq} = -\del^2 (\del^2 + \kappa^2)^2 \Bigl[(\del^2 + \tilu^2)^2 - B_1
\frac{\del^2_\sigma}{\del^2}(\del^2 + \tilu^2) + B_2 \del^2_\sigma
\Bigr]
  \comma
\eqe with \eqb
    B_1 = \lb \frac{2a_0 a_5 \kappa k}{\tilu} \rb^2 \comma \quad
    B_2 = \lb \frac{2a_0^2 u k}{\tilu} \rb^2 \period
\eqe

\msubsection{Fermionic fluctuation}

Let us move on to the fermionic part. To simplify the kinetic
term, we make an $SO(1,9)$ rotation with \eqb
   Q^a_{\ b} = \matrixii{q^a_{\ b}}{0}{0}{1_{7\times 7}}
     \comma \quad q^a_{\ b} = \frac{1}{M} \lb
                              \begin{array}{cc}
                                 a_0 \kappa & - W_s \\
                                 0 & M_s \\
                                 -W & a_0 \kappa W_s/W
                              \end{array}
                               \rb
 \comma
\eqe where $ a,b =(t,\phi_5,\rho, \theta, ....)$ and $s=(5,1)$.
After this rotation, one finds the following fluctuation operator,
\eqb
    \Dhat_F &=& \sigbar^i \del_i  + a \sigbar^{345}
      + c \sigbar^{023} + d \sigbar^{123} \comma
\eqe with \eqb
   a = a_0\kappa \frac{M_5}{M} \comma \quad c = \frac{a_5\kappa}{2MW} (a_0^2 k^2 + M^2)
   \comma \quad d = - \frac{ a_0^2 a_5 \kappa uk}{2MW} \period
\eqe Here, we have labelled $(t,\phi_5,\rho,\theta,\rho_4) $ as $
(0,1,3,4,5)$ and $(\phi_1,...)$ as $(2,6,7,8,9)$.

With the explicit forms of $\sigma^a$ in (\ref{sigmas}), $\Dhat_F$
takes a simple form \eqb
   \Dhat_F &=& \matrixii{\Delta^+_F}{0}{0}{\Delta^-_F} \otimes 1 \comma
\eqe where \eqb
   \Delta^+_F = \matrixii{\Delta^+_1}{0}{0}{\Delta^+_2} \comma \qquad
   \Delta^+_1 &=& \matrixii{\del_- + i(c+d)}{a}{-a}{\del_+ - i(c-d)} \comma
\eqe
 $ \Delta^+_2 (a,c,d) = \Delta^+_1(a,-c,-d) $, $
    \Delta^-_F (a,c,d)  =  \Delta^+_F (-a,c,d) $,
and $ \del_\pm = \del_\tau \pm \del_\sigma$. From these, one finds
that \eqb
  && {\rm pf} \, \Dhat_F = \det \Delta^+_F \det \Delta^-_F \comma \nn \\
  && \det \Delta^+_F  = \det \Delta^-_F  \\
     && \quad = \
   \Bigl[\del^2 + 2i (d \del_\tau + c \del_\sigma) + (c^2 -d^2 + a^2) \Bigr]
   \Bigl[\del^2  - 2i (d \del_\tau + c \del_\sigma) + (c^2 -d^2 + a^2) \Bigr]
  \period \nn
\eqe This gives the characteristic frequencies \eqb
   \omega_n = \ep_1 d + \ep_2 \sqrt{(n + \ep_1 c)^2 + a^2}
 \comma
\eqe with $\ep_1, \ep_2 = \pm 1$. Although the parameters
$(a,c,d)$ might look different at first from those in \cite{PTT},
one can show that they are actually the same (up to signs) by
using relations among parameters.

It turns out that we do not have to perform further
 rotations as in the previous cases. However, it is
useful to note that the terms with $i d \del_\tau$ can be removed
by a simple``rotation": \eqb
   e^{\pm id \tau} [\del^2 \pm 2i (d \del_\tau + c \del_\sigma)
    + (c^2 -d^2 + a^2)]    e^{\mp id\tau}
   = \del^2 + (a^2 + c^2) \pm 2i c \del_\sigma \period
\eqe

\par\bigskip
\msubsection{One-loop effective action}

To compute the one-loop effective action, we collect all
contributions, including the ghost part, and expand the operators
for large $\nu$, to find that \eqb
  i \Gamma^{(1)} & = & - \Tr \log \Bigl( 1 + \frac{b_1}{\del^2 + \nu^2} \Bigr)
     + 4 \Tr \log \Bigl( 1 + \frac{b_2 + 2ic \del_\sigma}{\del^2 + \nu^2} \Bigr)
     \label{SL2G} \\
         && \quad  \ -\half \Tr\log \lbb 1 + \frac{2b_3}{\del^2 + \nu^2}
           +  \frac{b_3^2 + B_2 \del_\sigma^2}{(\del^2 + \nu^2)^2}
             -B_1 \frac{\del^2_\sigma}{\del^2} \Bigl( \frac{1}{\del^2 + \nu^2}
  +  \frac{b_3}{(\del^2 + \nu^2)^2}\Bigr) \rbb \comma \nn
\eqe with \eqb
   && b_1 = \kappa^2 - \nu^2 \comma \quad
   b_2 = a^2 + c^2 - \nu^2 \comma \quad  b_3 = \tilu^2 - \nu^2  \period
\eqe Then, up to and including $\calO(1/\nu)$, the one-loop
effective action in this case becomes \eqb
   i \Gamma^{(1)} &\sim & + iTL M^2 \cdot C \comma \label{G1SL2}
\eqe where $ M^2 \sim \alpha(1+\alpha) k^2 $ and $C$ is given by
(\ref{Cpn}).

%
\msection{Correction to space-time energy}

Summarizing, up to and including $\calO(1/\nu) \sim \calO(1/J)$,
the one-loop effective actions are given by a universal form, \eqb
   i \Gamma^{(1)} &\sim & \mp iTL M^2 \cdot C \comma \label{G1}
\eqe with the minus sign for the $J_1=J_2$ three-spin and the
two-spin $SU(2)$ cases, and the plus sign  for the $SL(2)$ case.
We call the former two cases the $S^5$ case in the following. The
parameter $\nu$ was defined in (\ref{3spin2}) for the $S^5$ case,
and above (\ref{sl2largenu}) for the $SL(2)$ case, which was
related to the total $SO(6)$ spin by $J \sim \sql \nu$ for large
$\nu$. $T$ and $2\pi L$ were the volumes of the world-sheet
directions $\tau$ and $\sigma$, respectively, whereas $C$ was
given by (\ref{Cpn}). $M$ was given in terms of the winding
numbers by equations below (\ref{Q1}) and (\ref{eq1}). It also has
a unified geometrical expression
 $M^2 = \half \eta^{ij} G_{\mu\nu} \del_i x^\mu \del_j x^\nu
 = \sqrt{- \det h_{ij}} $
with $h_{ij}$ the induced metric. In terms of spins, this was
expressed  as
\eqb
  M^2 &=& \half \eta^{ij} G_{\mu\nu} \del_i x^\mu \del_j x^\nu \label{Msq} \\
  &\sim& \lmb \begin{array}{lll}
          {\displaystyle 2 m^2 J_1/(2J_1 + J_3) } & &  \ (J_1=J_2 \ \,
          \mbox{3-spin $$case}) \\
           \alpha(1-\alpha) (m_1-m_2)^2 \comma
            & {\displaystyle \alpha = J_1/(J_1+J_2)}  &  \ (SU(2) \ {\rm case}) \\
            \alpha(1+\alpha) k^2 \comma & {\displaystyle \alpha = S/J} & \ (SL(2) \ {\rm case})
         \end{array}
          \period \right. \nn
\eqe
Formally, the results in (\ref{G1}) for  the $SU(2)$ and the
$SL(2)$ cases are related by replacing the filling fractions as
$\alpha_{SU(2)} \to -\alpha_{SL(2)}$ with an identification of
$m_1-m_2$ and $k$.

\par\bigskip\ni
\msubsection{Comparison with the gauge theory side}

From these one-loop effective actions, one can obtain the
corrections to the space-time energy $E^{(1)}$. First, let us note
that, in general, a one-loop effective action is interpreted as
the one-loop correction to the (world-volume) energy with the
expectation values of fields fixed (see, e.g., \cite{Weinberg}). A
simple way to confirm this is to express the one-loop effective
action as a summation over characteristic frequencies (as in
(\ref{Gammasum})). In our case, this means that $\Gamma^{(1)}$ is
proportional to the one-loop correction to the world-sheet energy:
\eqb
    \Gamma^{(1)} = -T  E^{(1)}_{\rm 2d} \period
\eqe Second, the correction to the world-sheet energy is
translated into that of the space-time energy \cite{FT0} as \eqb
    E^{(1)} = \frac{1}{\kappa} E^{(1)}_{\rm 2d} \comma
\eqe because of $t = \kappa \tau$. Therefore, up to and including
$\calO(1/J^2)$, \eqb
      E^{(1)} \ \sim \ \pm C {M^{2} \o \kappa}  + \calO(1/J^3)
      \ \sim \ \pm \sql C {M^{2} \o J}
       \label{E1M2}
    \comma
\eqe with the plus sign for the $S^5$ case, and the minus sign for
the $SL(2)$ case. Here and in the following, we set $L=1$.
Comparing these with the classical expressions, \eqb
  E^{(0)} = \lmb \begin{array}{l} { \displaystyle J + \lambda \frac{M^{2}/2}{J}
     + \calO( \lambda^{2} /J^{3}) } \qquad \qquad (S^5 \ {\rm case}) \\
       {\displaystyle  S+J + \lambda \frac{M^{2}/2}{J}
     + \calO( \lambda^{2} /J^{3}) } \qquad (SL(2) \ {\rm case})
                 \end{array}
  \right. \comma
\eqe one finds that the $M^{2}$-dependence of $ E^{(1)}$ is the
same as that of the first subleading term in the classical
expression.

One can also compare (\ref{E1M2}) with the gauge theory results.
On the gauge theory side, the corresponding quantity is the $1/J$
correction to the anomalous dimension at order $\lambda$
\cite{BTZ}-\cite{HLPS}, \eqb
   \gamma^{(1)} = \pm \frac{\lambda}{2J^2}\lbb M^2 + K_{SL(2),SU(2)}(M^2) \rbb
     \comma \label{anomdim}
\eqe where $K_{SL(2),SU(2)}(M^2)$ are certain functions of $M^2$
and called the ``anomaly" terms \cite{BTZ,HLPS}. Thus, $E^{(1)}$
has the same $M^2$-dependence of the first (``zero-mode") term of
$\gamma^{(1)}$. Note that, if $C$(+ possible contributions from
finite counter terms) $\sim 1/\nu$, the $J$- and the
$\lambda$-dependences also match: $E^{(1)} \sim \lambda M^2/J^2$.
In addition, the zero-mode ($n=0$) part of $C$, i.e., $1/2\nu$,
gives exactly the same contribution to $E^{(1)}$ as the zero-mode
term of $\gamma^{(1)}$.

\par\bigskip\ni
\msubsection{Relation to the results in the literature}

In the literature, the one-loop correction to the space-time
energy
 has been studied  by summing up the characteristic
frequencies. It was computed numerically for the $J_1 = J_2$
three-spin case and the $SU(2)$ case with two equal spins in
\cite{FPT}, and for the $SL(2)$ case in \cite{PTT}. The result of
the $SL(2)$ case was matched with the $1/J$ correction to the
anomalous dimension (\ref{anomdim}) including the anomaly terms
\cite{BTZ}. An agreement was also found in the $SU(2)$ case with
two equal spins, up to subtleties due to the instability. Thus, in
these cases, \eqb
   E^{(1)}_{\rm literature} = \gamma^{(1)} \period
\eqe This is to be compared with our result (\ref{E1M2}). First,
the zero-mode parts of $E^{(1)}_{\rm literature}$ are the same as
those on the gauge theory side in (\ref{anomdim}) \cite{BTZ} and
hence as ours. However, the full expression of $E^{(1)}_{\rm
literature}$ and the numerical results in \cite{FPT} do not have a
simple polynomial dependence on the winding numbers or $M^2$. As
for the $J$-dependence, if we evaluate $C$ by a simple summation
explained below (\ref{Cpn}), $C \sim 1/2$  and the $J$-dependence
does not match, either. (However, see the discussions below
(\ref{Cpn}) and (\ref{anomdim}).) Thus, the results in the
literature and ours appear to be different. In this subsection, we
would like to discuss this point in more detail.

First, we note that a one-loop effective action may be in general
expressed either as a functional determinant (or a trace as in
(\ref{simGamma})), or as a summation over characteristic
frequencies (as in (\ref{Gammasum})). Thus, our method is
equivalent to those in \cite{FPT,PTT} before the summation. The
agreement of the zero-mode contributions supports this.
 The apparent difference of the results is then understood as due to
the difference of the methods of evaluation of the same quantity
and extraction of its large $J$ behavior.
We have used the effective
action language, since this is a useful tool for our purpose. In
terms of it, the point of our method may be summarized as a choice
of convenient functional bases for evaluating the one-loop
effective action and space-time energy.

In order to see how apparent parameter dependence looks different
depending on the way of evaluation and expansion, it may be useful
to consider
 a simplified example of a one-loop effective action,\footnote{For the purpose
 of this subsection, one may take
an even simpler example such as $ \Tr \log(\del^2 + \nu^2 + k^2) -
\Tr \log (\del^2 + \nu^2)$, although this is logarithmically
divergent. }
\eqb
  i\Gamma^{(1)} = \Tr \log(\del^2 + \nu^2) + \Tr \log(\del^2 + \nu^2 + 2k^2)
           -  2\Tr \log(\del^2 + \nu^2 + k^2)
           \comma \label{simGamma}
\eqe
with $\nu$ large and $k$ of order one. Following our
computations in the previous sections, one can combine all terms,
to find
\eqb
  i\Gamma^{(1)} = \Tr \log \Bigl( 1+\frac{2k^2}{\del^2 + \nu^2} \Bigr)
     - 2 \Tr\log \Bigl( 1+\frac{k^2}{\del^2 + \nu^2} \Bigr) \period
\eqe
The expansion of the logarithms gives a well-defined large
$\nu$ expansion. For example, the first non-trivial term is
\eqb
  i\Gamma^{(1)} &\sim& \Tr \frac{-k^4}{(\del^2+\nu^2)^2}  = - \frac{iT}{4}
    \sum_n \frac{k^4}{(p_n^2 + \nu^2)^{3/2}} \sim
    - i TL \frac{k^4}{4\nu^2} \cdot c \comma \label{simG11}
\eqe
where we have approximated the sum by an integral, and  $ c =
\int dx (x^2+1)^{-3/2} = 2$.

On the other hand, by a standard procedure, $\Gamma^{(1)}$ is
written as a sum over characteristic frequencies as
\eqb
  i\Gamma^{(1)} &\sim & iT \sum_n \lb \sqrt{p_n^2 + \nu^2}
    + \sqrt{p_n^2 + \nu^2 + 2k^2}
     - 2 \sqrt{p_n^2 + \nu^2 + k^2} \rb \comma \label{Gammasum}
\eqe
up to an additive constant. Following \cite{PTT,BTZ}, one may
expand each term with respect to $ 1/\nu^2$, to find
\eqb
   i\Gamma^{(1)} \sim iT \nu \sum_n \sum_{j=1} \frac{c_{j}(p_n,k)}{\nu^{2j}}
   \comma \label{simG12}
\eqe
with $c_j(p_n,k)$ $2j$-th polynomials of $p_n$ and $k$. The
parameter dependence of (\ref{simG12}) is in fact very different
from that of (\ref{simG11}).

Here, we would like to make a comment on the expansion in
(\ref{simG12}). Although the original series is convergent, $p_n$
in the summation can be larger than $\nu$, and the terms with
larger $j$ become potentially more divergent.\footnote{ In this
example, $c_1 = 0$, whereas, in the original case in
\cite{PTT,BTZ}, the first term in the expansion gives the finite
leading correction. Regarding the divergences for $c_{j\geq 2}$,
see a footnote in \cite{BTZ}.} This subtlety can be rephrased also
as follows. First, let us define
\eqb
   f(s) \equiv  \Gamma^{(1)}/(T\nu)
          = \sum_n \, g_n(s) \comma
\eqe
where $ s = 1/\nu^2$ and $g_n(s) = \sqrt{1+ s p_n^2}
  + \sqrt{1+ s(p_n^2 +2k^2)} - 2 \sqrt{1 + s(p_n^2 +k^2)}$.
The above expansion corresponds to the expansion of $f(s)$ around
$ s = 0 \, (J = \infty)$,
\eqb
   f(s) & = & f(0) + s f'(0) + \cdots \nn \\
    & \sim &  s \sum_n g'_n(0) \period
\eqe Note that, to obtain the last expression, one needs to use
$f'(0) = \sum_n g'_n(0)$. By a basic result of analysis, a
(sufficient) condition for such termwise differentiability is the
uniform convergence of $\sum_n g'_n(s)$ around $s=0$, in addition
to the pointwise convergence of $\sum_n g_n(s)$ there. However,
for large $n$, $g_n(s) \sim \sqrt{s} k^4/(p_n)^3$ and hence $
g'_n(s) \sim k^4/(\sqrt{s}p_n^3)$, which implies that $\sum_n
g'_n(s)$ is not uniformly convergent around $s=0$. Since larger
and larger $n$ becomes relevant as $s$ approaches zero, it seems
to be difficult to study this subtlety numerically.\footnote{
Although the characteristic frequencies take different and more
complicated forms, the expansion coefficients of $\Gamma^{(1)}$ in
\cite{PTT} are divergent as in (\ref{simG12}), except in the
leading term. Thus, the expansion is not defined. We also remark
that the order of limits, i.e., $ \lim_{N\to \infty} \sum_{n}^N$
first and then $ \nu \, (\sim J) \to \infty$, has been changed in
the expansions in \cite{PTT} and (\ref{simG12}). In the numerical
computation in \cite{FPT}, a summation of the form $S = \sum_n
f(n/\kappa)$ is first split into two parts as $S = (\sum_{|n| <
N+1} + \sum_{|n| > N+1}) f(n/\kappa) \equiv S_1 +S_2 $. $S_1$ is
then numerically evaluated, whereas $S_2$ is regarded as the
error. To estimate $S_2$, an approximation by an integral in
\cite{FT2} is used: $
   f(x_i) = \int_{x_i}^{x_{i+1}}dx \, g(x) + \calO(1/\kappa^5)   \comma
$
where $g(x)$ is a certain function, and $x_i = n_i/\kappa$. Although $S_2$ is estimated
in \cite{FPT} as
$ S_2 = \int_{N/\kappa}^{\infty} dx \, g(x) + \calO(1/\kappa^5)$, the above
approximation of $f(x_i)$ by an integral
is just for a single term $f(x_i)$ and thus the total error for $S_2$ derived from it
is $\calO(L/\kappa^5)$, where $L$ is the number of terms in $S_2$.
Since $L=\infty$ in this case, the estimation of the error should be refined.
In fact, the quantum corrections in Fig.1-5 in \cite{FPT} at $q= 0$
($\sg^2 =0$ in our notation) do not vanish, although it should vanish since $q=0$ corresponds
to the point-like BPS (BMN) solution.}

In addition, to consider the relation between expansion methods
and parameter dependence further, it may also be useful to observe
what would happen if we did not make the rotations of the
functional bases which we had performed to make the large $J$
expansion well-defined.  For example, in the $SL(2)$ case, if we
use $D_{pq}^1$ in (\ref{DBpq}) instead of $\Dhat_{pq}^1$ in
(\ref{DhatBpq}), we have
\eqb
  - \half \Tr \log \Bigl(1+ \frac{b'_1 \del^2_\tau + b'_2 \del^2_\sigma
    + b'_3 \del_\tau \del_\sigma }{\del^4 + 4 \nu^2 \del^2_\tau} \Bigr)
\eqe
instead of the second line in (\ref{SL2G}). Here, $b'_{1,2}$
are of order one, $b'_3$ is of order $\nu$, and we have used (\ref{d4d2}).
Suppose that one can expand the logarithm.
Then, evaluating the $\nu$-dependence similarly to section 4, one
finds that there are infinitely many terms up to and including
$\calO(1/\nu)$ due to the infrared behavior. Thus, this expansion
may not be valid in general. A formal expansion gives
\eqb
  i\Gamma^{(1)} \sim iT\frac{k^2}{\nu}
  [c'_1  + c'_2 (kL)^2 + c'_3 (kL)^4 \cdots \ ]
  \comma
\eqe
with $c'_j$ some constants. The $E^{(1)}$ which is read off
from this $\Gamma^{(1)}$ scales as $1/J^2$ and has complicated
$k^2$-dependence as in (\ref{anomdim}) with the anomaly terms.

We have discussed how the difference between our
results and those in the earlier works arises. Actually, our work
is an attempt to overcome the difficulties in the earlier works
discussed above.
In our method, the
one-loop effective action and space-time energy can be computed as
a systematic and well-defined expansion of $1/J$ including higher
order terms. It is also possible to obtain closed forms of the expansion
coefficients such as $C$ in (\ref{Cpn}).

As discussed at the end of
section 4, one has to regularize infinities in the intermediate
calculations, and the final finite part
should be fixed so as to maintain the symmetry of the theory.
(Note that this issue is common to the methods in the earlier works.)
In fact, one can confirm that our results maintain the symmetry:
the final result is compactly expressed as a geometrical invariant
(see (\ref{Msq})) and correctly vanishes in the BPS cases
($m, m_{1,2}, k = 0$). It is an interesting question if there
are further possibilities to shift the finite part while keeping
the symmetry maintained. Also, a very non-trivial
agreement of the string and the gauge theory results is found in \cite{BTZ}
based on \cite{PTT}, in spite of the difficulties in \cite{PTT}
concerning the large world-sheet momenta $n \gg J$.
Therefore,  the excitation modes with large $n$
would be important to understand the differences among our results,
those in \cite{FPT,PTT} and those in the gauge theory.
Note that such high excitation modes are not well-understood
in the context of AdS/CFT correspondence.
For example, it is not known to what these
correspond on the gauge theory side even in the pp-wave case.

\msection{Conclusions}

In this paper, we discussed quantum fluctuations of the constant
radii rotating strings in $AdS_5 \times S^5$. Using a functional
method, we developed a systematic method to compute the one-loop
sigma-model effective actions in closed forms as expansions for
large spins. A point was the change (rotation) of functional bases
to make the expansion well-defined. As examples, we explicitly
evaluated the leading terms (up to and including $\calO(1/J)$) for
the strings in the $SO(6)$ sector with two equal spins, the
$SU(2)$ sector, and the $SL(2)$ sector. We would like to note
that, in our method, it is straightforward to compute higher order
terms (up to any given order, in principle). We moreover obtained
the one-loop corrections to the space-time energy up to and
including $\calO(1/J^2)$. Comparing these with the finite size
corrections to the anomalous dimension on the gauge theory side,
we found that the dependence on the winding numbers and the
filling fractions agreed with that of the
``non-anomalous''(zero-mode) part on the gauge theory side.
Relation to the earlier results in the literature was also
discussed.

An obvious future direction is to probe quantum effects for more
general cases such as the folded and the circular solutions in
\cite{FT3}, by generalizing the method in this paper. In fact,
this was one of the original motivations of this work. Also, the
comparison with the results on the gauge theory side or in
\cite{PTT,BTZ} indicates that it is important to understand the
issue of the order of limits, as has been pointed out in the
literature of the rotating string/spin chain correspondence.

\vspace{6ex}
\begin{center}
  {\bf Acknowledgments}
\end{center}

We would like to thank N. Berkovits, D. Ghoshal,  N. Hatano, M.
Hatsuda, Y. Hikida, T. Mateos, K. Mohri, N. Ishibashi, Y. Okada,
S.-J. Rey, K. Sakai, M. Shiroishi,  B. Stefanski, Y. Susaki, Y.
Takayama, Z. Tsuboi, A. Yamaguchi, and K. Yoshida for useful
discussions and conversations. We would especially like to thank
A. Tseytlin for useful comments and discussions on an earlier
version of the manuscript in January, correcting errors in it, and
informing us of part of the results in \cite{BTZ} prior to
publication. H. F. and Y.S. are also grateful to California
Institute of Technology and Seoul National University,
respectively, for warm hospitality. The work of Y.S. was supported
in part by Grant-in-Aid for Young Scientists (B) No.\,16740124 and
Overseas Advanced Educational Research Practice Support Program
No.16-077 from the Japan Ministry of Education, Culture, Sports,
Science and Technology.

\vspace{6ex}
\setcounter{section}{0} \appsection{Evaluation of $C$}

In this appendix, we explain how to evaluate $C$ in (\ref{Cpn})
for large $\nu$. As discussed in section 4, the result can depend
on the way of calculation (regularization).

First, we evaluate $C$ by first summing the terms in the
parenthesis in (\ref{Cpn}) for given $n$ and then summing over
$n$. This sum can be approximated by an integral using the
Euler-Maclaurin formula, \eqb
  && \sum_{n=0}^N f(a+n \delta)  \\
  && \quad = \ \frac{1}{\delta} \int_a^{a+N\delta} f(x) \, dx
    + \half [f(a) +f(a+N\delta)]  +   \frac{\delta}{12} [f'(a+N\delta)-f'(a)]
   + \calO(\delta^2 f'')  \comma \nn
\eqe where the prime stands for the derivative. Setting $\delta =
1/(\nu L)$ and $f(x) =  \delta \cdot \bigl( \frac{x^2 +
1/2}{\sqrt{x^2 + 1}} - x \bigr)$, we then find that, for large
$N$, \eqb
   C \sim - \frac{\delta}{2} + 2  \sum_{n=0}^N f(n \delta)
     = + \half + \frac{1}{6(\nu L)^2}
      + \calO\Bigl( \frac{1}{(\nu L)^3}, \frac{\nu L}{N} \Bigr) \period
\eqe Since $f''(x)$ is bounded for $ x \in (0,\infty)$, the error
does not grow, as discussed in section 4.

Second, for example, we may also adopt the zeta-function
regularization (see, e.g., \cite{Voros}) for each divergent sum in
$C$. For this purpose, we introduce \eqb
  \zeta(z,\mu) \equiv \sum_{n \in \bfZ} \frac{1}{(n^2 +\mu^2)^z}
   = \frac{1}{\Gamma(z)} \int_0^\infty ds \, s^{z-1}
          e^{-\mu^2 s} \theta_3(is/\pi)
 \comma
\eqe where $\mu>0$ and $\theta_3(\tau) = \sum_{n \in \bfZ}
q^{n^2/2}$ with $q = e^{2\pi i \tau}$. The second equation gives
an analytic continuation in $z$. We then make a change of
variables, $u = \mu^2 s$, perform the modular transformation
$\theta_3(\tau) = (-i\tau)^{-1/2} \theta_3 (-1/\tau)$, and expand
the resultant theta function as $\theta_3(i\pi \mu^2/u) = 1 + 2
\sum_{n=1} e^{- (\pi \mu n)^2/u}$. After the integration, the
terms from the sum over $n \geq 1$ are expressed by a Bessel
function as $\frac{2\sqrt{\pi}}{\Gamma(z)}(\mu/\pi n)^{1/2 -z}
K_{1/2 -z}(2 \pi \mu n)$. These are exponentially suppressed for
large $\mu$, since $K_\nu(x) \sim x^{-1/2} e^{-x}$ for large $x$.
Thus, we find that \eqb
   \zeta(z,\mu) = \mu^{1-2z} B\bigl( \half,z- \half \bigr) + \calO(\mu^{-z}e^{-2\pi\mu})
\eqe for large $\mu$, where $B(x,y)$ is the beta function.
Applying this to each sum in $C$, we encounter some singular
expressions
 at intermediate steps. We regularize
them by shifting $z$ as $z + \epsilon$. Consequently, \eqb
  C
  \sim \frac{1}{(\nu L)^2}
    \Bigl[  \zeta\bigl( -\half+\epsilon,\nu L \bigr)
   - \frac{1}{2} (\nu L)^2  \zeta \bigl( \half + \epsilon,\nu L \bigr) - 2 \zeta(-1) \Bigr]
    \period
\eqe Carefully taking the limit $\epsilon \to 0$, we find that the
leading contributions from the first two terms cancel each other,
and \eqb
  C \sim - \half + \frac{1}{6(\nu L)^2}
   + \calO\Bigl( \frac{e^{-2\pi \nu L}}{\sqrt{\nu L}}, \epsilon \Bigr)
   \period
\eqe

%
\def\thebibliography#1{\list
 {[\arabic{enumi}]}{\settowidth\labelwidth{[#1]}\leftmargin\labelwidth
  \advance\leftmargin\labelsep
  \usecounter{enumi}}
  \def\newblock{\hskip .11em plus .33em minus .07em}
  \sloppy\clubpenalty4000\widowpenalty4000
  \sfcode`\.=1000\relax}
 \let\endthebibliography=\endlist
%
%
\vskip 10ex
\begin{center}
 {\bf References}
\end{center}
\par \smallskip

\end{document}